\documentclass[10pt,conference,letterpaper]{IEEEtran}
\usepackage[letterpaper,left=0.635in,right=0.64in,top=0.75in,bottom=1in]{geometry}
\usepackage[USenglish]{babel}
\usepackage{times,amsmath}
\usepackage[pdftex]{graphicx}
   \graphicspath{{./pdf/}{./png/}{./jpeg/}{./eps/}{./png_old/}}
   \DeclareGraphicsExtensions{.pdf,.png,.jpeg,.eps}
\usepackage{cite}
\usepackage{url}
\usepackage{fmtcount}
\usepackage{color}
\usepackage[tight,footnotesize]{subfigure}
\usepackage{multicol}
\usepackage{amssymb}
\usepackage{bm}
\usepackage{wasysym}
\usepackage{dblfloatfix} 
\usepackage{calc}
\usepackage{fix-cm}
\usepackage[activate={true,nocompatibility},final,tracking=true,factor=500,stretch=40,shrink=40]{microtype}
\usepackage{stmaryrd}
\usepackage{textcomp}
\usepackage{mathtools}
\usepackage[libertine]{newtxmath}
\usepackage{scalerel}
\usepackage{fancyhdr}
\def\cal{\fam2}

\newcommand{\e}{\mathrm{e}}
\newcommand{\rd}{{\rm d}}

\newcommand{\const}{{\rm const}}
\newcommand{\half}{{\frac{1}{2}}}
\newcommand{\Ibl}{{\llbracket}}
\newcommand{\Ibr}{{\rrbracket}}

\DeclareMathOperator\erfc{erfc}
\DeclareMathOperator\erf{erf}
\newcommand{\appropto}{\mathrel{\vcenter{
  \offinterlineskip\halign{\hfil$##$\cr
    \propto\cr\noalign{\kern2pt}\sim\cr\noalign{\kern-2pt}}}}}
\definecolor{DarkRed}{rgb}{0.5,0,0}
\definecolor{DarkGreen}{rgb}{0,0.3333,0}
\definecolor{DarkerGreen}{rgb}{0,0.3333,0}
\definecolor{DarkBlue}{rgb}{0,0,0.75}
\definecolor{RoyalBlue}{rgb}{0,0.1373,0.4000}
\definecolor{NavyBlue}{rgb}{0,0,0.5020}
\definecolor{CobaltBlue}{rgb}{0,0.2784,0.6706}
\definecolor{lightlightgray}{rgb}{0.96875,0.96875,0.96875}
\definecolor{cyan}{rgb}{0,1,1}
\newcommand{\beginlabel}[2]{%
\begin{#1}\label{#2}}
\begin{document}
\pagestyle{fancy}
\lhead{\small Nikitin and Davidchack}
\rhead{\small M-ary Aggregate Spread Pulse Modulation in LPWANs for IoT applications}
\title{M-ary Aggregate Spread Pulse Modulation\\ in LPWANs for IoT applications}
\author{\IEEEauthorblockN{Alexei V. Nikitin}
\IEEEauthorblockA{
Nonlinear LLC\\
Wamego, Kansas, USA\\
E-mail: avn@nonlinearcorp.com}
\and
\IEEEauthorblockN{Ruslan L. Davidchack}
\IEEEauthorblockA{Sch. of Mathematics and Actuarial Sci.,\\
U. of Leicester, Leicester, UK\\
E-mail: rld8@leicester.ac.uk}}
\maketitle
\begin{abstract}
In low-power wide-area networks (LPWANs), various trade-offs among the bandwidth, data rates, and energy per bit have different effects on the quality of service under different propagation conditions (e.g. fading and multipath), interference scenarios, multi-user requirements, and design constraints. Such compromises, and the manner in which they are implemented, further affect other technical aspects, such as system's computational complexity and power efficiency. At the same time, this difference in trade-offs also adds to the technical flexibility in addressing a broader range of IoT applications. This paper addresses a physical layer LPWAN approach based on the Aggregate Spread Pulse Modulation (ASPM) and provides a brief assessment of its properties in additive white Gaussian noise (AWGN) channel. In the binary ASPM the control of the quality of service is performed through the change in the spectral efficiency, i.e., the data rate at a given bandwidth. Implementing M-ary encoding in ASPM further enables controlling service quality through changing the energy per bit (in about an order of magnitude range) as an additional trade-off parameter. Such encoding is especially useful for improving the ASPM's energy per bit performance, thus increasing its range and overall energy efficiency, and making it more attractive for use in LPWANs for IoT applications.
\end{abstract}
\maketitle

\section{Introduction} \label{sec:introduction}
In the Aggregate Spread Pulse Modulation (ASPM)~\cite{Nikitin20pulsed,Nikitin21ASPMpatent} the information is encoded in the amplitudes~$A_j$ and/or the ``arrival times" $k_j$ of the pulses in a digital ``pulse train"~$\hat{x}[k]$ with only relatively small fraction of samples having non-zero values:
\beginlabel{equation}{eq:ptrain}
  \hat{x}[k] = \sum_j \Ibl k\!=\!k_j\Ibr\, A_j\,,\vspace*{-1mm}
\end{equation}
where $k_j$ is the sample index of the $j$-th pulse, $A_j$ is its amplitude, and the double square brackets denote the {\it Iverson bracket\/}~\cite{Knuth92two}
\vspace*{-1mm}
\beginlabel{equation}{eq:Iverson bracket}
  \Ibl P\Ibr  = \left\{
  \begin{array}{cc}
    \!\! 1 & \mathrm{if} \; P \; \mbox{is true}\\
    \!\! 0 & \mathrm{otherwise}
  \end{array}\right.,\vspace*{-1mm}
\end{equation}
where $P$ is a statement that can be true or false. The average ``pulse rate"~$f_\mathrm{p}$ in such a train is $f_\mathrm{p}=F_\mathrm{s}/N_\mathrm{p}$, where $F_\mathrm{s}$ is the sample rate, and $N_\mathrm{p}=\langle k_j-k_{j-1}\rangle$ is the average interpulse interval. Note that for $N_\mathrm{p}\gg 1$ the pulse rate is much smaller than the Nyquist rate. Also note that for ${N_\mathrm{p}\gg 1}$ this train has a large peak-to-average power ratio (PAPR) even when~$|A_j|=\const$, and is generally unsuitable for use as a modulating signal. However, the designed pulse train~$\hat{x}[k]$ given by~(\ref{eq:ptrain}) can be ``re-shaped" by linear filtering:
\vspace*{-1mm}
\beginlabel{equation}{eq:ptrain filtered}
   x[k] = (\hat{x}\ast \hat{g})[k] = \sum_j A_j\,\hat{g}[k\!-\!k_j]\,,\vspace*{-2mm}
\end{equation}
where $\hat{g}[k]$ is the impulse response of the filter and the asterisk denotes convolution. The filter~$\hat{g}[k]$ can be, for example, a lowpass filter with a given bandwidth~$B$. If the filter~$\hat{g}[k]$ has a sufficiently large time-bandwidth product (TBP)~\cite{Gabor45theory,Vetterli95wavelets}, most of the samples in the reshaped train~$x[k]$ will have non-zero values, and~$x[k]$ will have a much smaller PAPR than the designed sequence~$\hat{x}[k]$. Such low-PAPR signal can then be used for modulating a carrier. If the combination of the amplitude~$A_j$ and the arrival time~$k_j$ of a pulse provides~$M$ distinct ``states," each pulse can encode~$\log_2M$ bits, and the raw bit rate $f_\mathrm{b}$ in such a train is $f_\mathrm{b}=f_\mathrm{p}\log_2M$. When $B\gg f_\mathrm{b}\!=\!(F_\mathrm{s}/N_\mathrm{p})\log_2M$, it results in a low-rate message encoded in a wideband waveform.

For example, for the arrival times in~(\ref{eq:ptrain}) one can use
\vspace*{-1mm}
\beginlabel{equation}{eq:kj}
  k_j = jN_\mathrm{p} + \Delta{k}[m_j]\,,\vspace*{-1mm}
\end{equation}
where~$\Delta{k}$ is a positive integer, $0\le \Delta{k}[m_j] < N_\mathrm{p}$, and $\Delta{k}[m]\ne \Delta{k}[l]$ for~$m\ne l$. Then for $m_j=1,2,\dots,M$ and~$A_j=\const$ the pulse train given by~(\ref{eq:ptrain}) encodes $\log_2M$ bits per pulse. We will refer to such M-ary encoding with $A_j=\const$ as ``unipolar." Another bit can be added by using~$A_j=(-1)^{a_j}$, where $a_j$ is either ``0" or ``1," and we will refer to such signaling as ``bipolar." Then for bipolar M-ary signaling equation~(\ref{eq:ptrain}) can be rewritten as
\vspace*{-1mm}
\beginlabel{equation}{eq:ptrain Mary}
  \hat{x}[k] = \sum_j \Ibl k = jN_\mathrm{p} \!+\! \Delta{k}[m_j]\Ibr\, (-1)^{a_j}\,,\vspace*{-1mm}
\end{equation}
where $m_j=1,2,\dots,M/2$ and $a_j$ is either ``0" or ``1."

For a given designed pulse sequence~$\hat{x}[k]$ the spectral, temporal and amplitude structures of the reshaped train~$x[k]$ will be determined by the choice of~$\hat{g}[k]$. In particular, it may be desirable to select a filter~$\hat{g}[k]$ that minimizes the PAPR of~$x[k]$. Note that if the time duration of~$\hat{g}[k]$ extends over multiple interpulse intervals, the instantaneous amplitudes and/or phases~\cite{Picinbono97instantaneous} of the resulting waveform are no longer representative of individual pulses. Instead, they are a ``piled-up" aggregate of the contributions from multiple ``stretched" pulses. 

The key property of the large-TBP pulse shaping filter (PSF)~$\hat{g}[k]$ is that its autocorrelation function (ACF), i.e., the convolution of~$\hat{g}[k]$ with its matched filter~$g[k]=\hat{g}[-k]$, has a much smaller TBP, in particular, sufficiently smaller than the ratio~$B/f_\mathrm{p}$. Then, after demodulation and analog-to-digital (A/D) conversion in the receiver, the encoded binary sequence can be recovered by filtering with~$g[k]$ and sampling the resulting pulse train at $k = jN_\mathrm{p} \!+\! \Delta{k}[m]$ (i.e., using~$g[k]$ as a decimation filter).

A good choice for the PSF would be a pulse that combines a small TBP of its ACF (e.g., close to that of a Gaussian pulse) with ACF's compact frequency support. An example would be a raised-cosine (RC) filter~\cite[e.g]{Proakis06digital} with unity roll-off factor. The minimum required (Nyquist) sample rate for such a filter will be double its (baseband) physical bandwidth~$B$, and the sample rate~$F_\mathrm{s}$ can be expressed as ${F_\mathrm{s}=2N_\mathrm{s}B}$, where~$N_\mathrm{s}\ge1$ is the oversampling factor. To minimize the power consumption, the memory usage, and the computational complexity of the digital processing, it is beneficial to keep the sample rate in the transceivers designed for IoT applications as low as possible, i.e., to use $N_\mathrm{s}=1$. Through the rest of the paper, we will assume sampling with the Nyquist rate~${F_\mathrm{s}=2B}$.

Since for a given designed pulse sequence~$\hat{x}[k]$ the temporal and amplitude structures of the reshaped train~$x[k]$ are determined by the PSF~$\hat{g}[k]$, these structures can be substantially different even for the pulse shaping filters with the same ACF. As discussed in~\cite{Nikitin20pulsed}, one can construct a great variety of large-TBP pulse shaping filters $\hat{g}_i[k]$ with the same small-TBP ACF~$w[k]$, so that $(\hat{g}_i\ast g_i)[k]=w[k]$ for any~$i$, while the convolutions of any $\hat{g}_i(t)$ with $g_j(t)$ for $i\ne j$ (cross-correlations) have large TBPs. Further, this property will also effectively hold for the PSFs~$\hat{h}_i[k]$ such that~$\hat{h}_i[k]$ is the discrete Hilbert transform of~$\hat{g}_i[k]$, i.e., ${\hat{h}_i[k]=H\left\{ \hat{g}_i[k] \right\}}$~\cite{Bracewell2000FourierFULL, Todoran2008discrete}. Therefore, using various PSFs combinations we can design different coherent and noncoherent modulation schemes with emphasis on particular spectral and/or temporal properties of the modulated signal. 

\subsection{Binary (``one bit per pulse") encoding} \label{subsec:one bit}
For example, in~\cite{Nikitin21aggregate} we describe single-sideband, constant-envelope coherent and noncoherent ASPM configurations that use the ``equidistant" designed train
\vspace*{-.5mm}
\beginlabel{equation}{eq:ptrain equidistant}
  \hat{x}[k] = \sum_j \Ibl k\!=\!jN_\mathrm{p}\Ibr\, (-1)^{b_j}\vspace*{-1.5mm}
\end{equation}
to encode the binary sequence~$(b_1b_2\dots b_j\dots)$. The raw bit rate $f_\mathrm{b}$ in such a train is $f_\mathrm{b}=F_\mathrm{s}/N_\mathrm{p}$, where $F_\mathrm{s}$ is the sample rate and $N_\mathrm{p}$ is the number of samples between pulses. We further show that, predictably, for an additive white Gaussian noise (AWGN) channel the uncoded bit error rate (BER) $P_\mathrm{b}$ of these binary ASPM configurations can be expressed as
\vspace*{-1mm}
\beginlabel{equation}{eq:ASPM Pb}
  \begin{array}{cccc} \displaystyle
    \!\! P_\mathrm{b} = \half \erfc \left(\! \sqrt{\frac{E_\mathrm{b}}{N_0}} \right) = \half \erfc \left(\! \sqrt{\frac{N_\mathrm{p}\Gamma}{2}} \right) \quad \mbox{(coherent)}\\[4mm]
    \displaystyle
    \!\! P_\mathrm{b} = \half \exp \left(-\frac{E_\mathrm{b}}{2N_0}\right) = \half \exp \left(-\frac{N_\mathrm{p}\Gamma}{4}\right) \quad \mbox{(noncoherent)}
  \end{array}\!,\vspace*{-1mm}
\end{equation}
where $\erfc(x)$ is the complementary error function~\cite{Abramowitz72handbook}, $E_\mathrm{b}$~is the energy per bit, $N_0$~is the (one-sided) power spectral density of the noise, and $\Gamma$ denotes the signal-to-noise ratio (SNR) defined as ${\Gamma = (E_\mathrm{b}/N_0)\times(f_\mathrm{b}/B)}$. Thus, at a given bandwidth, in the binary ASPM the control of the quality of service is performed through the change in the interpulse interval~$N_\mathrm{p}$, i.e., the data rate.

\begin{figure*}[!t]
\centering{\includegraphics[width=11.8cm]{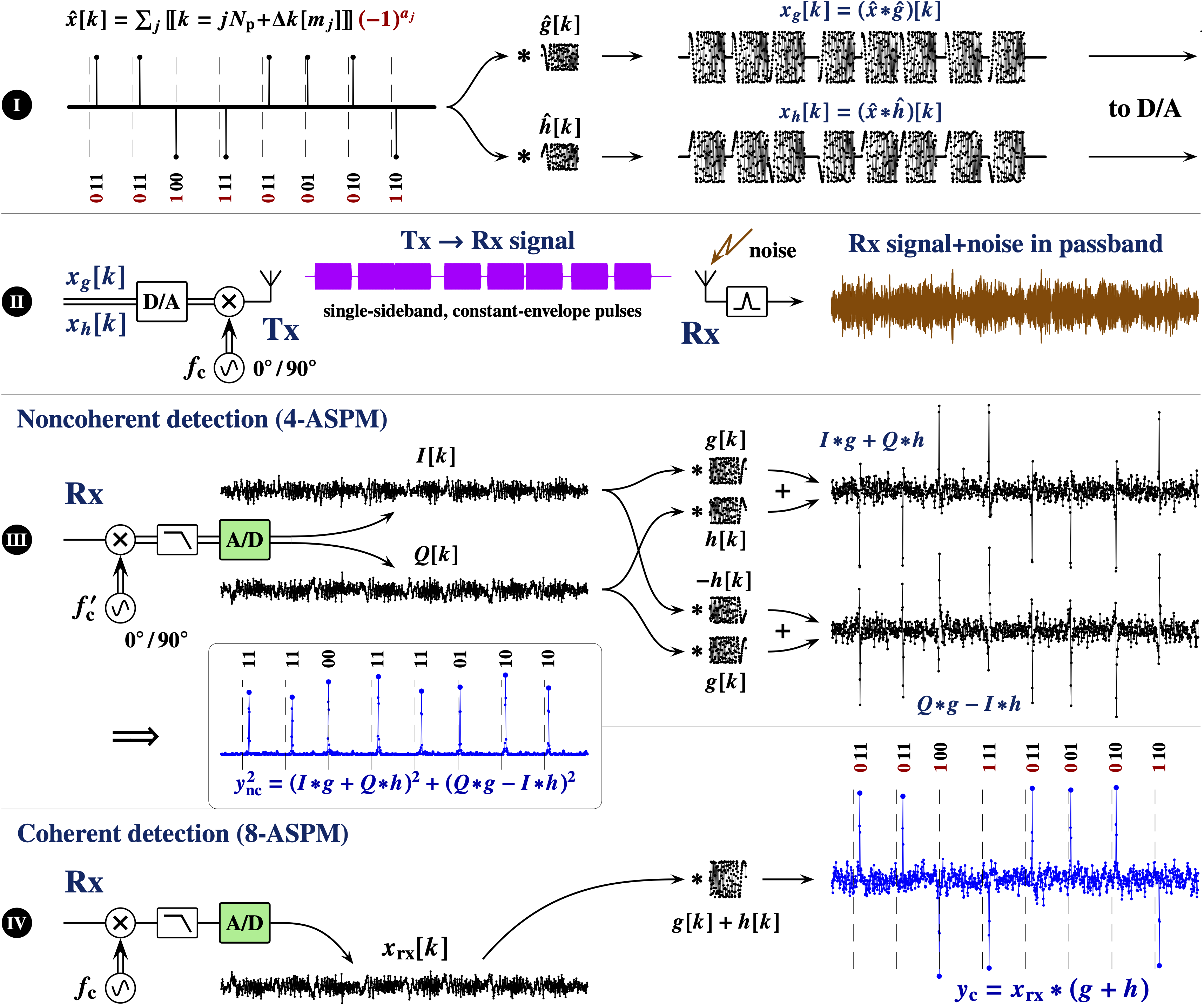}}
\caption{Illustration of single-sideband M-ary ASPM link with constant-envelope pulses and their noncoherent and coherent detection.
\label{fig:MaryASPM}}
\end{figure*}

\section{M-ary variants of ASPM} \label{sec:Mary}
In the binary ASPM, each pulse encodes one bit, hence the energy per bit~$E_\mathrm{b}$ and the energy per pulse~$E_\mathrm{p}$ are equal to each other, $E_\mathrm{b}=E_\mathrm{p}$. By encoding~$\log_2M$ bits per pulse with the same energy, the energy per bit is reduced to~$E_\mathrm{b}=E_\mathrm{p}/\log_2M$. Such encoding is especially useful for improving the ASPM's energy per bit performance, thus increasing its range and overall energy efficiency, and making it more attractive for use in LPWANs for IoT applications.

\subsection{Single-sideband M-ary ASPM with constant-envelope pulses} \label{subsec:Mary}
For example, Fig.~\ref{fig:MaryASPM} illustrates a single-sideband M-ary ASPM link which uses constant-envelope transmitted pulses and is suitable for both coherent and noncoherent detection.

In Fig.~\ref{fig:MaryASPM}(I), the designed pulse train~$\hat{x}[k]$ according to~(\ref{eq:ptrain Mary}) is filtered with~$\hat{g}[k]$ and~$\hat{h}[k]$ to form the shaped trains~$x_g[k]$ and~$x_h[k]$. After digital-to-analog (D/A) conversion, $x_g(t)$~and~$x_h(t)$ are used for quadrature amplitude modulation of a carrier with frequency~$f_\mathrm{c}$, providing the transmitted waveform ${x_g(t)\sin(2\pi f_\mathrm{c}t)} + {x_h(t)\cos(2\pi f_\mathrm{c}t)}$ (Fig.~\ref{fig:MaryASPM}(II)). If~$\hat{g}[k]$ and~$\hat{h}[k]$ are, say, the real and imaginary parts, respectively, of a nonlinear chirp with the desired ACF, e.g.
\vspace*{-1mm}
\beginlabel{equation}{eq:chirp}
  \hat{g}[k]+i\,\hat{h}[k] = \Ibl 0\!\le\! k\!<\!n\Ibr\, \exp \left( i\,\Phi[k]\right)\,,\vspace*{-1mm}
\end{equation}
where~$\Phi[k]$ is the phase, then this waveform will occupy only a single sideband with the physical bandwidth~$B$ equal to the baseband bandwidth of the chirp. In addition, if the pulses do not overlap (e.g., $N_\mathrm{p}>n+\max_m(\Delta{k}[m])$), this waveform will consist of constant-envelope pulses.

For noncoherent detection (Fig.~\ref{fig:MaryASPM}(III)), in the receiver's (Rx) quadrature demodulator the noisy passband signal is multiplied by the orthogonal sinusoidal signals from a local oscillator, lowpassed, and converted to the in-phase and quadrature digital signals $I[k]$ and $Q[k]$. Filtering $I[k]$ and $Q[k]$ with the pairs of the filters~$g[k]$ and~$h[k]$, as shown in Fig.~\ref{fig:MaryASPM}(III), produces the signal components $I\!\ast\! g + Q\!\ast\! h$ and $Q\!\ast\! g - I\!\ast\! h$. Further, the sum of squares of these components forms the unipolar pulse train $y^2_\mathrm{nc} = (I\!\ast\! g + Q\!\ast\! h)^2 + (Q\!\ast\! g - I\!\ast\! h)^2$ with the peaks corresponding to the pulses in the designed train~$\hat{x}[k]$. For coherent detection (Fig.~\ref{fig:MaryASPM}(IV)), after multiplication by~$\sin(2\pi f_\mathrm{c}t\!+\!\pi/4)$, lowpass filtering, and A/D conversion in the receiver, the resulting signal~$x_\mathrm{rx}[k]$ is filtered with ${g[k]\!+\!h[k]}$ to form the bipolar baseband pulse train $y_\mathrm{c} = x_\mathrm{rx} \ast (g+h)$ corresponding to the designed train~$\hat{x}[k]$.

Without loss of generality, the ACFs of~$\hat{g}[k]$ and~$\hat{h}[k]$ can be normalized to have the peak magnitudes equal to unity. Then, to avoid the interpulse interference in both coherent and noncoherent detection, we can require that
\vspace*{-1mm}
\beginlabel{equation}{eq:no interpulse}
  w[\Delta{k}[m] \!-\! \Delta{k}[l]] = v^2[\Delta{k}[m] \!-\! \Delta{k}[l]] = \Ibl m\!=\!l\Ibr\,,\vspace*{-1mm}
\end{equation}
where $w[k]=\half(\hat{g}\!\ast\! g + \hat{h}\!\ast\! h)$ and
\vspace*{-1mm}
\beginlabel{equation}{eq:noncoherent response}
  v^2[k] = \frac{1}{4}\, \left[ (\hat{g}\!\ast\! g + \hat{h}\!\ast\! h)^2 + (\hat{h}\!\ast\! g - \hat{g}\!\ast\! h)^2 \right].\vspace*{-1mm}
\end{equation}

Note that the A/D conversion in the ASPM receiver can be combined with intermittently nonlinear filtering described in~\cite{Nikitin19hidden, Nikitin19complementary}, to make the link robust to outlier interferences, e.g. impulsive noise commonly present in industrial environments~\cite{Courjault2020how}, and to increase the baseband SNR in the presence of such interferences. Since in the power-limited regime the channel capacity is proportional to the SNR, even relatively small increase in the latter will be beneficial.

\vspace*{-2mm}
\section{Uncoded BER performance of M-ary ASPM in AWGN channel} \label{sec:Mary BER}

\vspace*{-1mm}
\subsection{Noncoherent M-ASPM} \label{subsec:noncoherent M-ASPM BER}
Let us assume that we transmit the $j$-th pulse with~$m_j=1$, and in the receiver sample at $jN_\mathrm{p} + \{\Delta{k}[1], \Delta{k}[2], \dots, \Delta{k}[M]\}$. If $y^2_{m}=y^2_\mathrm{nc}\left[ jN_\mathrm{p} + \Delta{k}[m] \right]$, then the $j$-th symbol will be detected correctly when ${y^2_{1}>\max\{y^2_{2} ,y^2_{3}, \dots, y^2_{M}\}}$.

For AWGN with constant power density~$N_0$, and in the absence of interpulse interference, $Y^2_{m}$ for $m>1$ can be viewed as i.i.d. variables having chi-square distribution with $2$~degrees of freedom~\cite{Abramowitz72handbook}. Thus the cumulative distribution function of the random variable ${\cal{M}}=\max \left\{ Y^2_{2},Y^2_{3}, \dots, Y^2_{M} \right\}$ can be expressed as
\vspace*{-4mm}
\begin{equation} \label{eq:binom}
F_{\cal{M}}(x) = \left( 1\!-\!\e^{-\frac{x}{2}} \right)^{M-1} \!= \sum_{k=0}^{M-1} (-1)^k\binom{{M\!-\!1}}{k} \exp \left( -\frac{k}{2}x \right),\vspace*{-2mm}
\end{equation}
where~${\binom{n}{m} = \frac{n!}{(n-m)!\,m!}}$ is the binomial coefficient.

At the same time, $Y^2_{1}$ will have the noncentral chi-square distribution with $2$~degrees of freedom and the noncentrality parameter~$\lambda$ proportional to the peak power of the ``ideal" pulse~\cite{Abramowitz72handbook}, and its cumulative distribution function can be expressed as
\vspace*{-3mm}
\begin{equation} \label{eq:fY00}
F_{Y^2_{1}}(x) = 1-Q_1\left(\sqrt{\lambda},\sqrt{x}\right),\vspace*{-1mm}
\end{equation}
where $Q_1(a,b)$ is the Marcum $Q$-function defined as the integral
\vspace*{-2mm}
\begin{equation} \label{eq:Marcum}
Q_1(a,b) = \int_b^\infty\!\!\rd{x}\, x \exp \left( -\frac{x^2+a^2}{2} \right)\, I_0(ax)\vspace*{-1mm}
\end{equation}
for $a, b \ge 0$, and where $I_0(x)$ is the modified Bessel function of the first kind~\cite{Simon2002Nutall}.
Therefore, the symbol error probability~$P_\mathrm{s}$ can be expressed as
\vspace*{-2mm}
\begin{align} \label{eq:ASPM SER mbit}
P_\mathrm{s}(\lambda) &= P\left( Y^2_{1}\!<\! {\cal{M}} \right) = \int_0^\infty\!\!\!\rd{x}\, F_{Y^2_{1}}(x)\, \frac{\rd}{\rd{x}} F_{\cal{M}}(x)\nonumber\\
&= 1+\int_0^\infty\!\!\!\rd{x}\, F_{\cal{M}}(x)\, \frac{\rd}{\rd{x}} Q_1\left(\sqrt{\lambda},\sqrt{x}\right).\vspace*{-2.5mm}
\end{align}
Evaluating the integral in the right-hand side of~(\ref{eq:ASPM SER mbit}) by parts (see the Appendix), and noticing that the bit error probability~$P_\mathrm{b}$ is related to the symbol error probability~$P_\mathrm{s}$ as
\vspace*{-1mm}
\begin{equation} \label{eq:BER vs SER mbit}
P_\mathrm{b}(\lambda) = \frac{M}{2(M\!-\!1)}\, P_\mathrm{s}(\lambda)\,,\vspace*{-1mm}
\end{equation}
leads to the following expression for~$P_\mathrm{b}(\lambda)$ of noncoherent M-ASPM:
\vspace*{-3mm}
\begin{equation} \label{eq:ASPM BER binom}
P_\mathrm{b}(\lambda) = \frac{1}{2(M\!-\!1)} \sum_{k=2}^M (-1)^k\binom{M}{k}\, \exp \left( -\frac{k\!-\!1}{2k}\lambda \right).\vspace*{-1mm}
\end{equation}

\begin{figure}[!t]
\centering{\includegraphics[width=8.4cm]{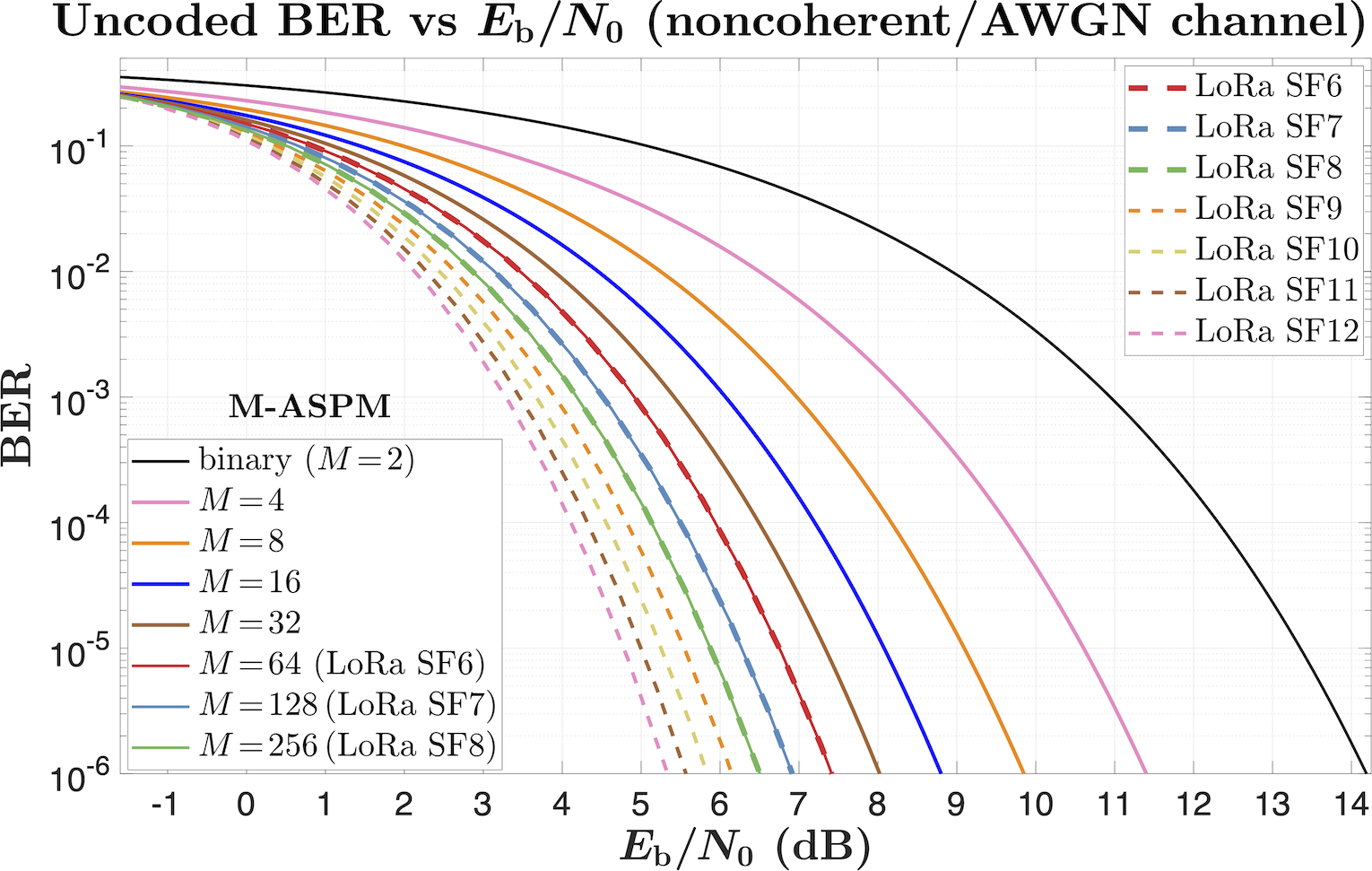}}
\caption{Uncoded BER vs $E_\mathrm{b}/N_0$ performances of LoRa (dashed lines) and single-sideband M-ASPM (solid lines) for noncoherent detection in AWGN channel.
\label{fig:BER EbN0 noncoherent}}
\end{figure}
\begin{figure}[!b]
\centering{\includegraphics[width=8.4cm]{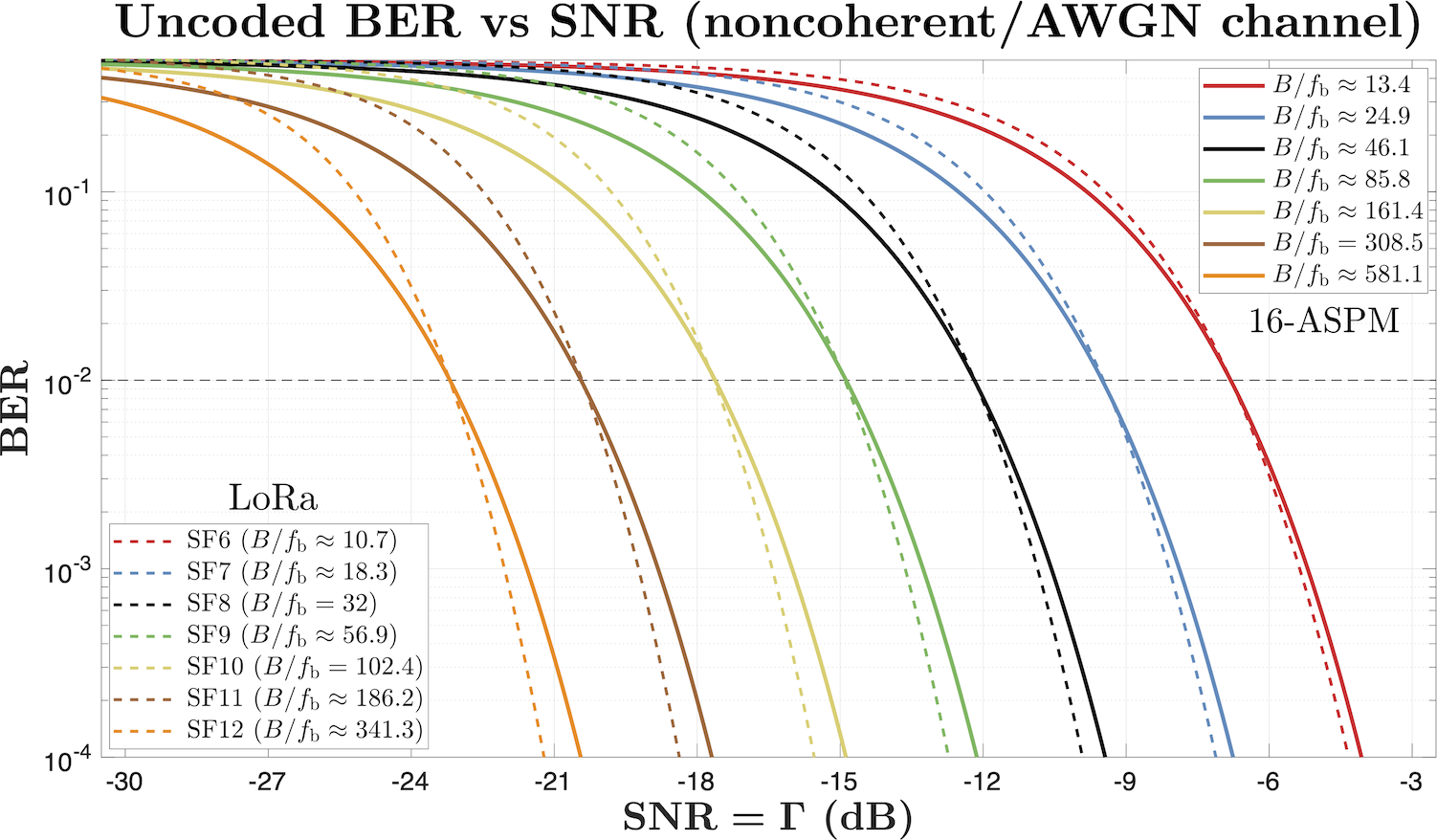}}
\caption{Uncoded BER vs SNR performances of LoRa (dashed lines) and single-sideband 16-ASPM (solid lines) for noncoherent detection in AWGN channel. For 16-ASPM, $B/f_\mathrm{b} = N_\mathrm{p}/8$.
\label{fig:BER SNR noncoherent}}
\end{figure}

\subsubsection{Value of noncentrality parameter~$\lambda$} \label{subsubsec:lambda}
The noncentrality parameter~$\lambda$ is the ratio of the baseband peak signal power~$A^2$ and the noise power~$\sigma^2_\mathrm{n}$, $\lambda=A^2/\sigma^2_\mathrm{n}$, and it can be expressed in several different ways, for example as
\vspace*{-1mm}
\begin{equation} \label{eq:lambda two}
  \lambda = \frac{2 E_\mathrm{b}}{N_0}\log_2M = \frac{2\sigma^2_\mathrm{c}}{N_0 f_\mathrm{b}}\log_2M =  2N_\mathrm{p}\frac{\sigma^2_\mathrm{c}}{N_0 F_\mathrm{s}} = N_\mathrm{p}\Gamma\,,\vspace*{-1mm}
\end{equation}
where~$\sigma^2_\mathrm{c}$ is the power of the modulated carrier, thus describing the service quality in terms of different physical and numerical parameters of the link. In~(\ref{eq:lambda two}), as before, the SNR is defined as ${\Gamma = (E_\mathrm{b}/N_0)\times(f_\mathrm{b}/B)}$. Note that the spreading factor in the M-ASPM is $B/f_\mathrm{b} = N_\mathrm{p}/(2\log_2M)$. Then, for example, in terms of the energy per bit $\gamma_\mathrm{b} = E_\mathrm{b}/N_0$, the bit error probability of noncoherent M-ASPM is
\vspace*{-1mm}
\begin{equation} \label{eq:ASPM BER binom EbN0}
P_\mathrm{b}(\gamma_\mathrm{b}) = \frac{1}{2(M\!-\!1)} \sum_{k=2}^M (-1)^k\binom{M}{k}\, \exp \left( -\frac{k\!-\!1}{k}\,\gamma_\mathrm{b}\, \log_2 M \right).\vspace*{-1mm}
\end{equation}

Note that, for a given~$\gamma_\mathrm{b}$, the bit error probability is a decreasing function of~$M$ and, for~${M\ge 64}$, is the same as the bit error probability of noncoherent LoRa with the spreading factor~$\mathrm{SF}=\log_2 M$~\cite{Baruffa20error}. This is illustrated in Fig.~\ref{fig:BER EbN0 noncoherent}, where the M-ASPM BER performance is compared with the respective performance of the noncoherent LoRa with different spreading factors. For LoRa, the BER approximation proposed in~\cite{Baruffa20error} is used, which is expressed as the product of the union bound on the bit error probability and a correction function.

In Fig.~\ref{fig:BER SNR noncoherent}, the BER vs. SNR performance of the noncoherent 16-ASPM ($B/f_\mathrm{b} = N_\mathrm{p}/8$) is compared with the respective performances of the noncoherent LoRa with different spreading factors. As can be seen from the figure, in terms of the energy per bit performance for uncoded $\mathrm{BER}=10^{-2}$ in an AWGN channel, the noncoherent 16-ASPM is approximately 60\% to 80\% as efficient as LoRa with the spreading factors ranging from~$\mathrm{SF}=6$ to~$\mathrm{SF}=12$.

\subsection{$E_\mathrm{b}/N_0$ efficiency of coherent M-ASPM} \label{subsec:coherent M-ASPM BER}
By using additional $M/2$~distinct pulse locations in the binary coherent ASPM, each pulse can encode $m=\log_2 M$ bits. For example, for~$M=16$, the pulse train
\vspace*{-.5mm}
\beginlabel{equation}{eq:ptrain four bipolar}
  \hat{x}[k] = \sum_j \Ibl k\!=\!jN_\mathrm{p} + (4a_j\!+\!2b_j\!+\!c_j)n\Ibr\, (-1)^{d_j},\vspace*{-2mm}
\end{equation}
where~$n$ is a nonzero integer, encodes a 4-bit sequence~$(a_1b_1c_1d_1\, a_2b_2c_2d_2 \dots a_jb_jc_jd_j\dots)$. To correctly identify a symbol in such M-ASPM, we need to correctly detect both the arrival time and the polarity of the pulse.

When the arrival time of a pulse with the peak magnitude~$|A|$ is known, the probability of correctly detecting the polarity of this pulse in the presence of AWGN with zero mean and variance~${\sigma_\mathrm{n}}^2$ can be expressed, using the complementary error function, as ${\half\erfc(-\mu)}$, where ${\mu = |A|/(\sigma_\mathrm{n}\sqrt{2})}$. We can further assume that $n$~in~(\ref{eq:ptrain four bipolar}) is sufficiently large, and thus interpulse interference is negligible (e.g. $n\ge 2$ for coherent detection and pulse shaping with the ACF as an RC pulse with unity roll-off factor). Then, for a pulse train with the peak magnitude of the pulses equal to~$|A|$, and $m=\log_2 M$ bits per pulse encoding, the bit error probability can be expressed as
\vspace*{-1mm}
\begin{equation} \label{eq:M-ASPM BER coherent}
P_\mathrm{b}(\mu) = \frac{M}{2(M\!-\!1)} \left[1- \half\erfc(-\mu)\,  P\left(|X_1|>\cal{M}\right) \right],\vspace*{-1mm}
\end{equation}
where~$X_1$ is a normal random variable with mean~$\mu\propto |A|$ and variance~$1/2$, and
\vspace*{-2mm}
\begin{equation} \label{eq:maxM}
{\cal{M}} = \max \left\{ |X_2|,|X_3|,\dots,|X_{\frac{M}{2}}| \right\},\vspace*{-1mm}
\end{equation}
where~$X_i$, ${i=2,3,\dots,M/2}$, are i.i.d. normal variables with zero mean and variance~$1/2$.

For $Y=|X_1|$, its cumulative distribution function is that of the folded normal distribution, which can be expressed as
\vspace*{-.5mm}
\begin{equation} \label{eq:folded normal}
F_Y \left(x;\mu \right) = \half \left[ \erf\left( x+\mu \right) + \erf\left( x-\mu \right) \right]\vspace*{-.5mm}
\end{equation}
for~$x\ge 0$. Then the probability to correctly detect the arrival time of the pulse is
\vspace*{-1mm}
\begin{align} \label{eq:P arrival}
&P\left(|X_1|>\cal{M}\right) = \int_0^\infty\!\!\!\!\rd{x}\, \left[F_Y \left(x;0\right)\right]^{\frac{M}{2}-1}\, \frac{\rd}{\rd{x}} F_Y \left(x;\mu \right)\nonumber\\
&= \int_0^\infty\!\!\!\!\rd{x} \left[\erf(x)\right]^{\frac{M}{2}-1} \left\{ \frac{1}{\sqrt{\pi}} \left[ \e^{-\left( x+\mu \right)^2} + \e^{-\left( x-\mu \right)^2} \right] \right\}.\vspace*{-1mm}
\end{align} 
For $\mu=0$ the right-hand-side integral is equal to~$2/M$, and for $\mu>0$ it can be easily evaluated numerically.

\begin{figure}[!t]
\centering{\includegraphics[width=8.4cm]{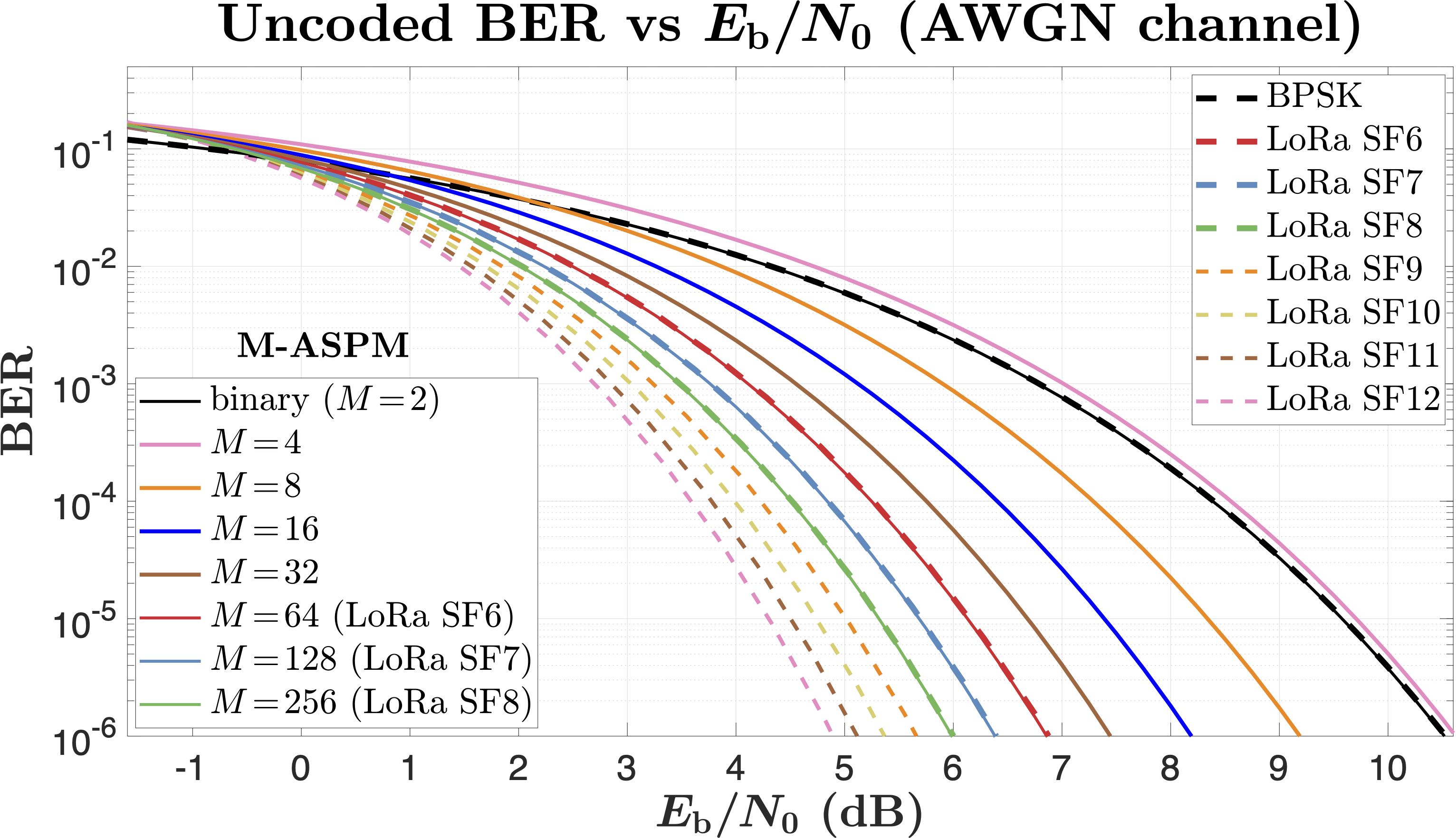}}
\caption{Uncoded BER vs $E_\mathrm{b}/N_0$ performances of BPSK and LoRa (dashed lines), and single-sideband M-ASPM (solid lines) for coherent detection in AWGN channel. For~${M\ge 64}$, M-ASPM $E_\mathrm{b}/N_0$ efficiency is that of LoRa with spreading factor~$\mathrm{SF}=\log_2 M$.
\label{fig:BER EbN0 coherent}}
\end{figure}
\begin{figure}[!b]
\centering{\includegraphics[width=8.4cm]{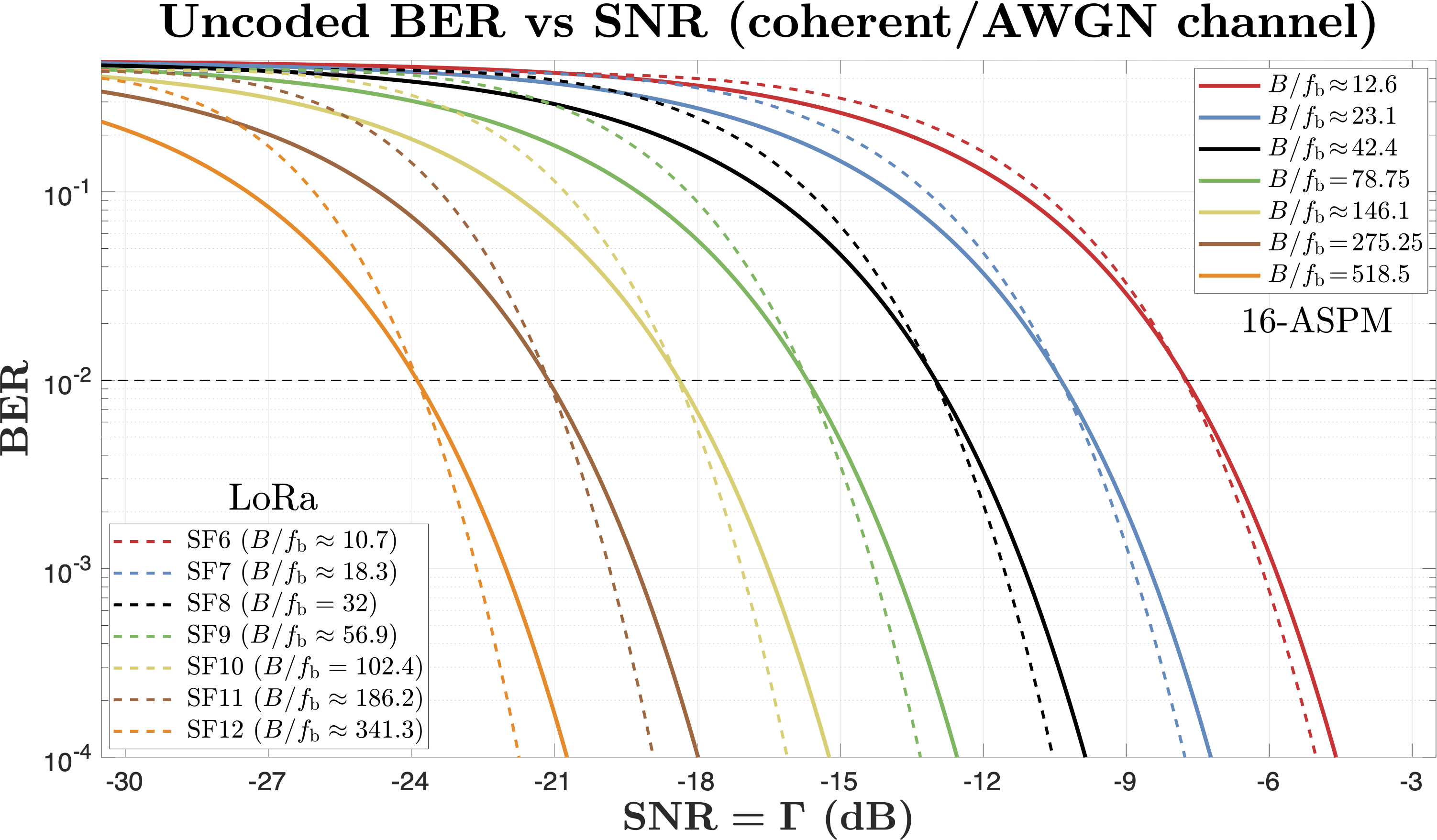}}
\caption{Uncoded BER vs SNR performances of LoRa (dashed lines) and single-sideband 16-ASPM (solid lines) for coherent detection in AWGN channel. For 16-ASPM, $B/f_\mathrm{b} = N_\mathrm{p}/8$.
\label{fig:BER SNR coherent}}
\end{figure}

\subsubsection{Value of~$\mu$} \label{subsubsection:mu}
For coherent detection, the ratio of the baseband peak signal power~$A^2$ and the noise power~$\sigma^2_\mathrm{n}$ is the same as for noncoherent detection~\cite{Nikitin21aggregate}, and thus ${\mu = |A|/(\sigma_\mathrm{n}\sqrt{2}) = \sqrt{\lambda/2}}$, where $\lambda$ is the noncentrality parameter of the noncoherent ASPM given by~(\ref{eq:lambda two}). Then, for example,
\vspace*{-1mm}
\begin{equation} \label{eq:mu}
  \mu = \sqrt{\frac{E_\mathrm{b}}{N_0}\log_2M}_{{}_{}} = \sqrt{\frac{N_\mathrm{p}\Gamma}{2}}_{{}_{}}\,,
\end{equation}
where ${\Gamma = (E_\mathrm{b}/N_0)\times(f_\mathrm{b}/B)}$ is the SNR. The bit rate $f_\mathrm{b}$ is related to the pulse rate $f_\mathrm{p}$ as $f_\mathrm{b} = f_\mathrm{p}\log_2{M}$, and, as before, the spreading factor in the M-ASPM is $B/f_\mathrm{b} = N_\mathrm{p}/(2\log_2M)$.

Fig.~\ref{fig:BER EbN0 coherent} shows computed uncoded BER vs $E_\mathrm{b}/N_0$ performances of BPSK and LoRa (dashed lines), and single-sideband M-ASPM (solid lines) for coherent detection in AWGN channel. Note that, just like in the noncoherent case, for~${M\ge 64}$ the M-ASPM $E_\mathrm{b}/N_0$ efficiency equals that of LoRa with the spreading factor~$\mathrm{SF}=\log_2 M$~\cite{Baruffa20error}.

In Fig.~\ref{fig:BER SNR coherent}, the BER vs. SNR performance of the coherent 16-ASPM ($B/f_\mathrm{b} = N_\mathrm{p}/8$) is compared with the respective performances of the coherent LoRa with different spreading factors. As can be seen from the figure, in terms of the energy per bit performance for uncoded $\mathrm{BER}=10^{-2}$ in an AWGN channel, the coherent 16-ASPM is approximately 70\% to 90\% as efficient as LoRa with the spreading factors ranging from~$\mathrm{SF}=6$ to~$\mathrm{SF}=12$.

\begin{figure}[!b]
\centering{\includegraphics[width=8.4cm]{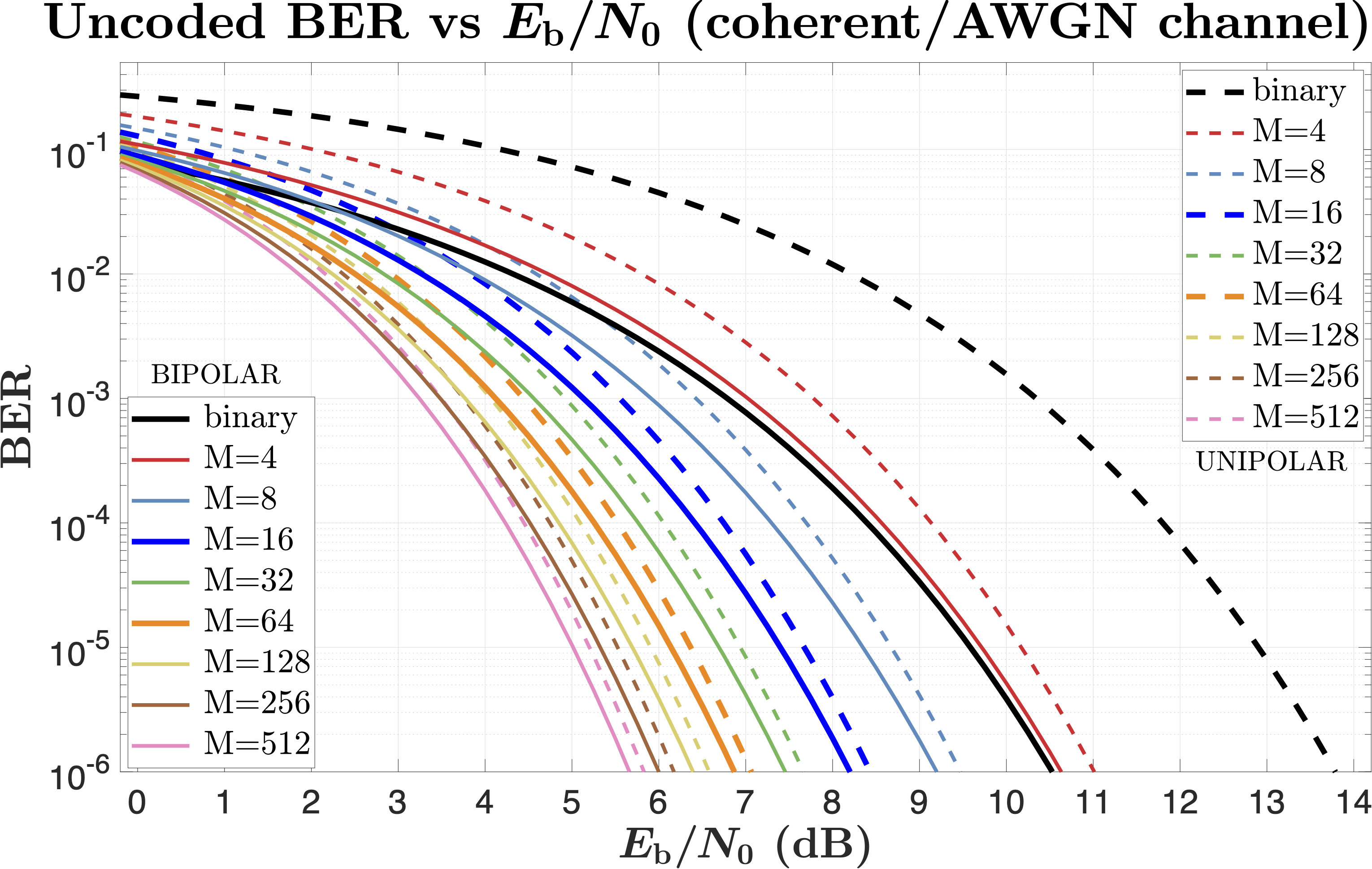}}
\caption{Comparison of uncoded AWGN BER vs $E_\mathrm{b}/N_0$ performances of unipolar (dashed lines) and bipolar (solid lines) signaling in coherent M-ASPM.
\label{fig:BER EbN0 coherent unipolar}}
\end{figure}

\subsubsection{Unipolar signaling} \label{subsubsec:unipolar}
For both noncoherent and coherent M-ASPM, bipolar encoding requires only $M/2$~distinct pulse locations. In comparison with the unipolar signaling, this doubles the maximum achievable data rate for a given~$M$. However, the noncoherent detection always requires obtaining $M$~samples per pulse, and has identical energy per bit and computational efficiencies for either bipolar or unipolar encoding. In contrast, after the synchronization has been obtained, the detection for bipolar encoding in coherent M-ASPM requires sampling at only~$M/2$ data points per pulse, thus halving the per-bit computational intensity of numerical processing. In addition, the bipolar signaling in coherent M-ASPM has the advantage of higher AWGN energy per bit efficiency compared to the unipolar signaling.

Indeed, the bit error probability for the unipolar coherent detection can be expressed as
\vspace*{-1mm}
\begin{align} \label{eq:Pb unipolar}
&P^\mathrm{up}_\mathrm{b}(\mu) = \frac{M}{2(M\!-\!1)} \left\{1- \int_0^\infty\!\!\!\!\rd{x}\, \left[\erf(x)\right]^{M-1}\, \frac{\rd}{\rd{x}} F_Y \left(x;\mu \right) \right\}\\
&= \frac{M}{4} \int_0^\infty\!\!\!\!\rd{x}\, \left\{ \frac{2}{\sqrt{\pi}}\, \e^{-x^2} \right\}\left[\erf(x)\right]^{M-2} \left[ \erf(x\!+\!\mu)+\erf(x\!-\!\mu) \right],\nonumber\vspace*{-2mm}
\end{align} 
and it can be shown that $P^\mathrm{up}_\mathrm{b}(\mu;M)>P_\mathrm{b}(\mu;M)$ for $\mu>0$. This is illustrated in Fig.~\ref{fig:BER EbN0 coherent unipolar}, that compares the uncoded AWGN BER vs. $E_\mathrm{b}/N_0$ performances of unipolar (dashed lines) and bipolar (solid lines) signaling in coherent M-ASPM. Predictably, the difference between $P^\mathrm{up}_\mathrm{b}(\mu;M)$ and $P_\mathrm{b}(\mu;M)$ becomes negligible in the limit of large~$M$.

\section{Simulated BER vs SNR performance of coherent and noncoherent 16-ASPM} \label{sec:simulation}
\begin{figure}[!b]
\vspace*{-3mm}
\centering{\includegraphics[width=8.4cm]{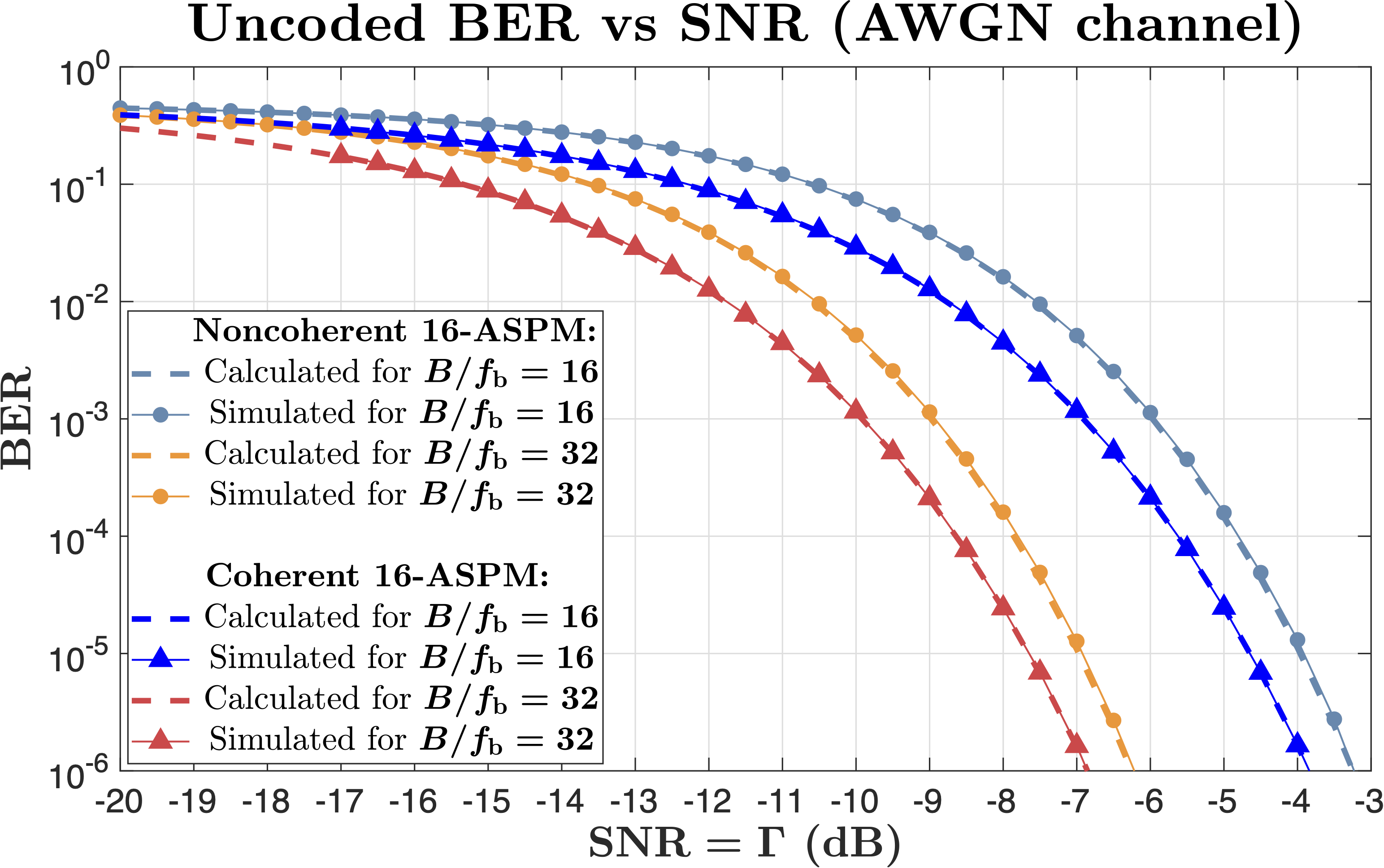}}
\caption{Calculated and simulated BERs as functions of AWGN SNRs for both coherent and noncoherent 16-ASPM links with $N_\mathrm{p} = 128$ and  $N_\mathrm{p} = 256$.{\tiny\newline~}
\label{fig:BER vs SNR}}
\end{figure}

Fig.~\ref{fig:BER vs SNR} compares the calculated (dashed lines) and the simulated (markers connected by solid lines) BERs for both coherent and noncoherent 16-ASPM links with the spreading factors ${B/f_\mathrm{b} = N_\mathrm{p}/8 = 16}$ and ${B/f_\mathrm{b} = N_\mathrm{p}/8 = 32}$.

For the coherent 16-ASPM, the designed pulse train~$\hat{x}[k]$ is given by~(\ref{eq:ptrain four bipolar}), where $n=2$. For the noncoherent 16-ASPM, the designed pulse train is
\vspace*{-1mm}
\beginlabel{equation}{eq:ptrain four unipolar}
  \hat{x}[k] = \sum_j \Ibl k\!=\!jN_\mathrm{p} + (8a_j\!+4b_j\!+\!2c_j\!+\!d_j)n\Ibr\,,\vspace*{-1mm}
\end{equation}
where $n=4$. In the transmitter, filtering~$\hat{x}[k]$ with the PSF~$\hat{g}[k]$ forms the modulating component~$x_\mathrm{I}[k]$, and filtering~$\hat{x}[k]$ with the PSF~$\hat{h}[k]$ forms the modulating component~$x_\mathrm{Q}[k]$. The ACF of~$\hat{g}[k]$ is an RC pulse with unity roll-off factor. The filter~$\hat{h}[k]$ approximates the discrete Hilbert transform of~$\hat{g}[k]$, i.e., ${\hat{h}[k]\approx H\left\{ \hat{g}[k] \right\}}$~\cite{Bracewell2000FourierFULL, Todoran2008discrete}, and thus $x_\mathrm{Q}[k]$~approximates the discrete Hilbert transform of~$x_\mathrm{I}[k]$, i.e., ${x_\mathrm{Q}[k]\approx H\left\{ x_\mathrm{I}[k] \right\}}$. Therefore, if after digital-to-analog conversion $x_\mathrm{I}(t)$~and~$x_\mathrm{Q}(t)$ are used for quadrature amplitude modulation of a carrier with frequency~$f_\mathrm{c}$, the resulting modulated waveform ${x_\mathrm{I}(t)\sin(2\pi f_\mathrm{c}t)} + {x_\mathrm{Q}(t)\cos(2\pi f_\mathrm{c}t)}$ occupies only a single sideband with the physical bandwidth~$B$ equal to the baseband bandwidth of~$\hat{g}[k]$.

In the coherent receiver, the noisy passband signal is multiplied by the signal~$\sin(2\pi f_\mathrm{c}t+\pi/4)$ from the local oscillator, lowpassed, and A/D converted to form the digital signal~$x_\mathrm{rx}[k]$, which is then filtered with ${g[k] + h[k]}$ to form the baseband pulse train
\vspace*{-1mm}
\beginlabel{equation}{eq:Rx coherent}
  y_\mathrm{c} = x_\mathrm{rx} \ast ( g\!+\!h )\,.\vspace*{-1mm}
\end{equation}
For noncoherent detection, in the receiver's quadrature demodulator the noisy passband signal is multiplied by $\sin(2\pi f_\mathrm{c}t+\varphi)$ and $\cos(2\pi f_\mathrm{c}t+\varphi)$, lowpassed, and A/D converted to the in-phase (I) and quadrature (Q) digital signals $I[k]$ and $Q[k]$. Then the received unipolar pulse train is formed as
\vspace*{-1mm}
\beginlabel{equation}{eq:Rx noncoherent}
  y^2_\mathrm{nc} = (I\!\ast\! g + Q\!\ast\! h)^2 + (Q\!\ast\! g - I\!\ast\! h)^2\,.\vspace*{-1mm}
\end{equation}

In the simulations, the bit error rates are determined by comparing the bit sequences extracted from the ``ideal" transmitted signals (without noise), and from the transmitted signals affected by AWGN with a given power spectral density~$N_0$.

\section{Conclusion} \label{sec:conclusion}
In this paper, we demonstrate how M-ary signaling setup can be utilized within the ASPM to improve its energy per bit performance, increasing its range and overall energy efficiency and making it more attractive for use in LPWANs for IoT. Using various combinations of pulse shaping filters in ASPM we can design numerous coherent and noncoherent modulations schemes, with emphasis on particular spectral and/or temporal properties of the modulated signal. This allows us to accommodate different propagation conditions in different IoT environments, meet diverse multiuser and physical layer security requirements, and, overall, add to the technical flexibility in addressing a broader range of IoT applications, both static and mobile.

\appendix
\renewcommand{\theequation}{\thesection.\arabic{equation}}
\setcounter{equation}{0}
Substituting~(\ref{eq:binom}) into~(\ref{eq:ASPM SER mbit}), and considering a single term in the sum:
\begin{subequations} \label{eq:Ps binom}
\begin{align}
  &\int_0^\infty\!\!\!\rd{x}\, \exp \left( -\frac{k}{2}x \right)\, \frac{\rd}{\rd{x}} Q_1\left(\sqrt{\lambda},\sqrt{x}\right) \tag{\ref{eq:Ps binom}}\\
  &= -\frac{1}{2} \int_0^\infty\!\!\!\rd{x}\, \exp\left( -\frac{\lambda+(k\!+\!1)\,x}{2} \right)\, I_0\left( \sqrt{\lambda x} \right) \label{eq:Ps binom a}\\
  &= -\frac{1}{k\!+\!1}\, \exp\left( -\frac{k}{2(k\!+\!1)} \lambda \right)\,, \nonumber
\end{align}
\end{subequations}
where we have used the the equalities~\cite{Simon2002Nutall,Brychkov2012some}
\begin{equation} \label{eq:dQdx}
\frac{\rd}{\rd{x}} Q_1\left(\sqrt{\lambda},\sqrt{x}\right) = -\half \exp\left( -\frac{\lambda+x}{2} \right)\, I_0\left( \sqrt{\lambda x} \right)\,,
\end{equation}
${Q_1(a,0) = 1}$, and ${Q_1(a,\infty) = 0}$, the substitution ${x=v^2/(k+1)}$ in~(\ref{eq:Ps binom a}), and the definition of the Marcum $Q$-function~(\ref{eq:Marcum}).
Further,
\begin{align} \label{eq:Ps final}
 P_\mathrm{s}(\lambda) &= 1-\sum_{k=0}^{M\!-\!1} \frac{(-1)^k}{k\!+\!1} \binom{M\!-\!1}{k}\, \exp \left( -\frac{k}{2(k\!+\!1)}\lambda \right) \nonumber\\
  &= \frac{1}{M} \sum_{k=2}^M (-1)^k\binom{M}{k}\, \exp \left( -\frac{k\!-\!1}{2k}\lambda \right).
\end{align}


\end{document}